\documentclass[aps,amsfonts,preprint,nofootinbib,superscriptaddress]{revtex4}

\def\S2{\bar{S}}

\def\and{a_{n}^\dagger}

\def\sn2d{\Sn2^\dagger}

\def\({\left(}
\def\){\right)}
\def\<{\left\langle}
\def\>{\right\rangle}

\newcommand\ee{\end{eqnarray}}      
\newcommand\be{\begin{eqnarray}}
\newcommand\ba{\begin{array}}           
\newcommand\ea{\end{array}}
\newcommand\eeq{\end{equation}}     
\newcommand\beq{\begin{equation}}
\usepackage{caption}

\usepackage{graphicx}

\begin{document}
\title{Time dependent Entanglement Entropy in  SYK models and Page Curve.}

\author{Daniel L. Nedel}
\email{daniel.nedel@unila.edu.br}
\affiliation{Universidade Federal da Integra\c{c}\~{a}o Latino-Americana, Instituto Latino-Americano de Ci\^{e}ncias da Vida e da Natureza, Av. Tancredo Neves 6731 bloco 06, CEP: 85867-970, Foz do Igua\c{c}u, PR, Brasil}

\begin{abstract}
In this work new interaction terms between two SYK models are proposed, which allow to define an interaction picture such that it is possible to calculate exactly the vacuum state's time evolution. It is shown that the vacuum evolves as a time dependent $SU(2)$ squezed state. The time dependent entanglement entropy is calculated and it has the same form of the Page curve of Black Hole formation and evaporation.
\end{abstract}

\maketitle

\section{Introduction}

One of the most interesting debates in theoretical physics in the last decades involves black hole formation and evaporation \cite{Haw1}. The great question is whether the process of formation and evaporation of a black hole can be described in a unitary fashion, thus compatible with quantum mechanics \cite{Haw2}. The $AdS_{d+1}/CFT_{d}$ correspondence has been playing an important role in this debate. In this scenario, the bulk quantum gravity theory in $d+1$ spacetime dimensions is dual to an ordinary d-dimensional conformal field theory that lives on the asymptotic boundary of the bulk spacetime \cite{malda1}. At first, the unitarity of the conformal field theory ensures that any quantum gravity phenomenon, including the formation and evaporation of a black hole, is unitary\footnote{Clearly, this is not the end of the debate. See references \cite{Almheiri:2020cfm,Penington:2019npb} for a more detailed discussion.}. Then the von Neumann entropy of the Hawking radiation should initially rise to a maximum value but then fall back down when the black hole evaporates. The system starts in a pure state and ends in a pure state, following the so-called “Page curve” \cite{Page1,Page2}.

An important case of the $AdS_{d+1}/CFT_{d}$ correspondence is the $d=1$ case, where the $CFT_1$ is interpreted as a conformal quantum mechanics.   Although the pure gravity in two dimensions is not well defined, the $AdS_2$ geometry appears as the near horizon limit of four or five dimensional extremal black holes  \cite{malda2,Kunduri:2007vf,Astefanesei:2006dd,Astefanesei:2007bf}. This implies that the microscopic explanation of Bekenstein-Hawking entropy of the extremal black holes is expected to be directly related to the $AdS_2/CFT_1$ correspondence \cite{Strominger:1996sh, Callan:1996dv, Strominger:1998yg}. An important issue was given in \cite{Azeyanagi:2007bj}, where it is presented an evidence that there are two systems of conformal quantum mechanics
(CQM) on the boundaries of the $AdS_2$ and that they are entangled with each other. It is shown that the black hole entropy is exactly the same as the
entanglement entropy of the two CQM .  

Recently, it has been argued that the fermionic degrees of freedom play an important role in calculating the black hole entropy \cite{Sachdev:2015efa}.  In fact, it was claimed in \cite{Iorio} that the Bekenstein bound itself descends from the Pauli principle. In the $AdS_2/CFT_1$ context, an important fermionic quantum mechanical model that helps to understand holographic entanglement entropy is the Sachdev-Ye-Kitaev (SYK) model, which consists of N Majorana fermions with random interaction \cite{Sachdev,Kitaev}.
When it is considered an interaction between two SYK models living in the boundaries of the $ADS_2$, the system is dual to a traversable wormhole \cite{Maldacena:2018lmt}. The coupling of two SYK models can be done using a teleportation protocol and it was shown that the coupled SYK models have an interesting phase diagram at finite temperature, displaying the usual Hawking-Page transition between the thermal AdS phase at low
temperature and the black hole phase at high temperature \cite{Sahoo,Maldacena:2018lmt,Gao:2019nyj,{Alet:2020ehp}}. The entanglement entropy between two SYK systems with bilinear coupling was calculated in \cite{Chen:2019qqe} and, in particular, it was shown that the ground state of the coupled system is close to a thermofield double state with particular temperature . 

In the present letter, we are going to show that a modification of the coupling between the two systems allows to write the evolution of the ground state as a time dependent $SU(2)$ entangled state. The vacuum evolution is obtained in an interaction picture where the bilinear interaction between the theories is  the states' time evolution operator. In this case, it is shown that the vacuum evolves to a state of maximum entanglement and returns to the initial pure state. The time dependent von Neumann entropy is calculated and the result reproduces the Page curve. Although the time to reach the maximum entanglement depends on the coupling between the theories, the maximum value of the time dependent entropy depends only on the Hilbert space dimension ($\mbox{dim} {\cal H}$) and it is written as $S_{max} = \ln(\mbox{dim} {\cal H})$. A more evolved situation is achieved when the initial state is prepared as a thermal vacuum using the approach of Thermo Field Dynamics (TFD).  It is shown that the evolution has the same characteristics. Now, from the point of view of one boundary, the system evolves from an entangled state, passes through a point of maximum entanglement and returns to the same entangled (thermal) state. However, in this case, a numerical analysis shows that the entanglement entropy only reproduces the Page curve at the large N limit. The time dependent state that is obtained via the temporal evolution of the thermal vacuum is similar to a non equilibrium time dependent TFD state. This state is achieved through TFD's typical Bogoliubov transformation, written in terms of time dependent fermionic oscillators. In the present work the focus will be in the tunneling Hamiltonian and vacuum evolution. The consequences of the new interaction terms in the general dynamics of the SYK model and the correlation functions will be not studied here. However, it will be shown that the dynamics of the vacuum in the interaction picture defined here is extremely rich and helps to understand the role of fermionic systems in black hole formation and evaporation.

\section{The model} 

Let's start by briefly reviewing the SYK model. A detailed discussion of the SYK model and its physical properties can be found in \cite{Gu:2019jub}. The most general complex SYK model can be described by the Hamiltonian

\begin{equation}
H= \sum_{i,j,k,l=1}^{N} J_{ij;kl} c^{\dagger}_{i} c^{\dagger}_{j} c_{k} c_{l} 
\label{eq:cSYK_Hamiltonian}
\end{equation} 

\noindent where the $c_{i}$, $c_i^\dagger$ are $N$ Majorana fermionic operators satisfying $\{ c_i, c_j^\dagger \} = \delta_{ij}$. The coefficients $J_{ij;kl}$ are complex Gaussian random numbers with 

\begin{equation}
    \overline{J_{ij;kl}} = 0 \quad , \quad \overline{ |J_{ij;kl}|^2 } = \frac{J^2}{8 N^3}
\end{equation}

\noindent and satisfy the symmetry constraints

\begin{equation}
J_{ij;kl}=-J_{ji;kl} = -J_{ij;lk} = J_{lk;ji}^{\ast}
\end{equation}

\noindent The main difference between the complex and the original model based on real Majorana fermions is that the complex one has a global $U(1)$ symmetry $c_i\rightarrow e^{\phi_i}c_i$. In ref. \cite{Sahoo}, the coupling of two SYK models is studied by analyzing the following Hamiltonian:

\begin{equation} 
H_{\kappa} = \sum_{ij;kl} J_{ij;kl} \sum_{a=1,2} c_{ia}^{\dagger}c_{ja}^{\dagger} c_{ka} c_{la} -  \mu\sum_{i }c^{\dagger}_{ia} c_{ia} + K
\label{eq:model_tunneling}
\end{equation} 

\noindent where the coupling constants $J_{ij;kl}$ are identical in systems 1 and 2, $\mu$ is related to chemical potential and it is set to zero here (it is not important for the vacuum dynamics) \cite{Chen:2018wvb}. The last term is

\begin{equation}
 K= \sum_{n} \kappa \left( e^{i \phi} c_{n1}^{\dagger} c_{n2} +  e^{-i \phi} c_{n2}^{\dagger} c_{n1} \right)
\end{equation}

\noindent where $\kappa$ and $\phi$ are real parameters. The K term is known as tunneling Hamiltonian. Now, two modifications are going to be made in the Hamiltonian (\ref{eq:model_tunneling}). Choosing $\phi=\frac{\pi}{2}$, the first one is just to perform the following Bogoliubov transformation
\be
c_{2n}(\theta)=b_n =c_{2n}^{\dagger}\cos(\theta_n) +c_{n1}\sin(\theta_n)
\ee

\noindent where $\theta_n$ is just a parameter that can be absorbed in the coupling constant of the theories. From now on it will be used $a_n$ and $b_n$ to represent the $c_{1n}$ and $c_{2n}(\theta)$ fermions. The Bogoliubov transformed tunneling Hamiltonian is

\begin{equation}
 K(\theta)= \sum_{n} i\kappa \cos(\theta_n)\left( a_nb_n-a_n^\dagger b_n^{\dagger} \right)\label{kt}
\end{equation}

\noindent By making the identifications

\begin{eqnarray}
\kappa\cos(\theta_n)=\lambda_n
\end{eqnarray}

\noindent the model that is going to be worked here is achieved. The Hamiltonian is $H=H_J+H_{\gamma}$, where

\begin{equation}
H_{\gamma}= \sum_{n=1}^{N}\left[\gamma_n(a_nb_n-a_n^\dagger b_n^{\dagger}\right ]
\label{Hevo}
\end{equation}

\noindent and $H_J$ is the part of the Hamiltonian that depends on $J_{ij;kl}$. The second modification is made just in $H_J$. A new interaction between fields $a$ and $b$ is introduced, such that $H_J=H_J+H_J'$, where, for $\theta=0$, 

\be
H_J &=&\sum_{i,j,k,l=1}^{N} J_{ij;kl}\left( a^{\dagger}_{i} a^{\dagger}_{j} a_{k} a_{l}+b^{\dagger}_{i} b^{\dagger}_{j} b_{k} b_{l}\right) \nonumber\\
H_J' &=&\sum_{i,j,k,l=1}^{N} J_{ij;kl}'\left(a_ia_jb_kb_l+ a^{\dagger}_ia^{\dagger}_jb^{\dagger}_kb^{\dagger}_l\right).  
\label{hnovo}
\ee

\noindent The second line of equation (\ref{hnovo}) is the new term and breaks the $U(1)$ symmetry of the original model\footnote{The $U(1)$ symmetry is already spontaneously broken in the original model, as shown in \cite{Klebanov:2020kck}}. Its importance will become clear below. Let us write $H_{Ja} =J_{ij;kl}\left( a^{\dagger}_{i} a^{\dagger}_{j} a_{k} a_{l}\right)$ and $H_{Jb}$ will be the same for $b$ fields. The Hamiltonians satisfy the following commutation relations:

\be
\left[H_{Ja},\:H_{\gamma}\right]&=& 2iJ_{ijkl}\left(a^{\dagger}_ia_ka_lb_j-a^{\dagger}_ia^{\dagger}_ja_kb^{\dagger}_l\right),\:\:\:
\left[H_{Jb},\:H_{\gamma}\right]=2iJ_{ijkl}\left(b^{\dagger}_ib_ka_jb_l-b^{\dagger}_ib^{\dagger}_ja^{\dagger}_lb_k\right) \nonumber \\ 
\left[H_J',\:H_{\gamma}\right] &=&2i\sum_{i,j,k,l=1}^{N}J_{ijkl}'\left[a^{\dagger}_ia_ja_kb_l+ a^{\dagger}_jb^{\dagger}_kb^{\dagger}_lb_i-b^{\dagger}_ia_jb_lb_k-a^{\dagger}_ia^{\dagger}_jb^{\dagger}_ka_l\right].
\label{comut}
\ee
 
 \noindent We can choose the constants $J_{ij;kl}'$ to be $J_{ij;kl}'=-J_{ij;kl},\:\:\:J_{ij;kl}'=-J_{kl;ij}$, so that
 
\begin{equation}
[H_J,H_{\gamma}]= 0. 
\label{comutH}
\end{equation}

\noindent These commutators play an important role in subsequent calculations. The vacuum is defined to be

\be
a_n\left|0\right\rangle_a&=&0\:\:\:n=1...N\\
b_n\left|0\right\rangle_b&=&0\:\:\:n=1...N
\ee

\noindent where $\left|0\right\rangle_A=\left|0_1,...0_N\right\rangle_A$ and the same for $\left|0\right\rangle_B$. Now, if it is chosen $\theta_n= 0, \forall n$, the vacuum of the original ($c_1, c_2$) and the Bogoliubov transformed system are related in the following way

\be
\left|0\right\rangle_a & = &\left|0\right\rangle_1 \nonumber \\
\left|0\right\rangle_b & = & \left|1_1 \, 1_2 \, \ldots \, 1_N\right\rangle_{2} .\label{vacuo}
\ee

\noindent The dimension of the  Hilbert space of one theory is $2^N$ and it will be shown that the maximum entropy is related to it.

\section{The $SU(2)$ entangled state and the von Neumann entropy}

Having established the fermionic model, we will study the evolution of the vacuum in this section. Note that, in general, the initial system is prepared to be a thermal state or the state of TFD at infinite temperature, which is the ground state of the Hamiltonian (\ref{kt}). The initial state will be the vacuum $\left |0\right\rangle=\left|0\right\rangle_{a}\otimes\left|0\right\rangle_{b}$, which corresponds to the TFD state at zero temperature. In order to define the vacuum's time evolution, the following interaction representation is defined for any state $\left|\alpha\right\rangle$ and operator $A$

\be
\left|\alpha,t\right\rangle_I&=& e^{itH_J}\left|\alpha,t\right\rangle_S \nonumber \\
A^I&=& e^{itH_J}A_Se^{-itH_J}
\ee

\noindent where $S$ stands for the usual Schrödinger representation and $I$ is the interaction representation. This implies that

\be
i\frac{\partial}{\partial t}\left|\alpha,t\right\rangle_I=H_{\lambda}^I\left|\alpha,t\right\rangle_I,\:\:\:
\frac{dA^I}{dt}=\frac{1}{i}[A_I,H_J]
\ee

\noindent so the operators have the evolution determined by $H_J^I$ and the states by $H_{\lambda}^I$. Now, owing to (\ref{comutH}), we get

\be
H_{\lambda}^I&=& e^{itH_J}H_{\lambda}e^{-itH_J}= H_{\lambda}
\ee 

\noindent and the vacuum state's time evolution is given just by $H_{\lambda}$

\begin{eqnarray}
\left|0(t)\right\rangle_I&=&e^{-it H_{\lambda}}\left|0\right\rangle_A\otimes \left|0\right\rangle_B \nonumber \\
&=&\prod_{k}\frac{\delta_{kk}}{\cos\left({\gamma_n t}\right)}\exp\left[\tan\left(\gamma_n t\right)a^{\dagger}_{k}b^{\dagger}_{k}\right]|0\rangle_A\otimes|0\rangle_B \ . \label{res1b}
\end{eqnarray}

Note that $\left|0(t)\right\rangle$ is a $SU(2)$ entangled state and it is similar to the TFD thermal vacuum. Choosing $\theta_n =0$, we set $\gamma$ to be the same for all oscillators, $\gamma_n= \gamma$. The system evolves from a pure state at $t = 0$, passes through a state of maximum entanglement at $t=\frac{\pi}{4\lambda}$  and, at $t = \frac{\pi}{2\lambda}$, it reaches a zero norm state. We can interpret this as another pure state, the null state; this is confirmed in the entropy calculation. Then a new cycle begins. We will restrict $\gamma t$ to $\gamma t \in [0,\frac{\pi}{2}]$.
  
  The maximum entanglement state corresponds to infinite temperature TFD state. As an example, let us write the maximum entanglement state for $N=1$. Using (\ref{vacuo}), the time dependent vacuum at $t=\frac{\pi}{4\lambda}$ for $N=1$ is
  
\begin{equation}
\left|0(t=\frac{\pi}{4\lambda})\right\rangle=\frac{1}{\sqrt{2}}\left(\left|0\right\rangle_1\otimes\left|1\right\rangle_2+ \left|1\right\rangle_1\otimes\left|0\right\rangle_2\right)
\end{equation}

\noindent which is just the Bell state.
  
  Let us define the density matrix

\begin{equation}
\rho(t)= |0(t)\rangle\langle 0(t)| \,.
\end{equation}

\noindent The  reduced density matrix is calculated by tracing over the $B$ degrees of freedom:

\begin{eqnarray}
 \rho _A &=&T{r_{\{ B\} }} \rho \\ \nonumber
 &=& \sum\limits_{\{ {\ell _i}\} } {\sum\limits_{\{ {{\tilde \ell }_i}\} }  {\langle \{ {\ell _i}\} |\langle \{ {{\tilde \ell }_i}\} |\rho \left| {\left\{ {{\ell _i}} \right\}} \right\rangle |\{ {{\tilde \ell }_i}\} \rangle } } \,,\\ \nonumber
&=&\prod_{k=1}^{N}\frac{1}{\cos(\gamma t)}\sum_{n_k}\left[\tan^2(\gamma t)\right]^{n_k}|n_k\rangle\langle n_k| 
\label{dmatrix}
\end{eqnarray}

 \noindent where  $\{n_i\}=\{n_i\}^N_{i=1}=n_1,\dots,n_N$ and we have used the notation $\tilde{n}_i$ to represent $b$ states. If we had taken the trace on system $A$, we would have the same result, as it must be for a bipartite system like this one. 

Let us calculate the von Neumann entropy relative to this reduced density matrix:

\begin{eqnarray}
S_A(t)&=& -Tr \rho _A \log \rho _A \nonumber \\
&=&  -\sum_{n=1}^{N}[\sin^2\gamma t\ln ( \sin^2\gamma t)
+\cos^2\gamma t\ln (\cos^2\gamma t)] \,, . 
\label{timeentropy}
\end{eqnarray}      

\begin{figure}[!h]
    \includegraphics[width=.5\textwidth]{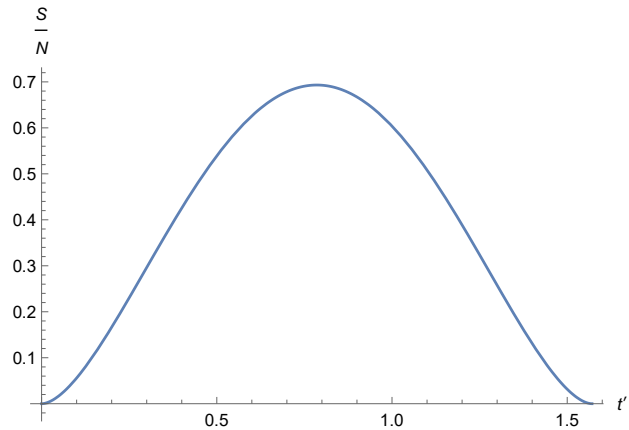}
    \caption{The behavior of $\frac{S(t)}{N}$  as a $t'=\gamma t$ function, from Eq.(\ref{timeentropy}). }
    \label{entro2}
\end{figure}

\noindent The entropy resembles the Page curve of black hole entropy, as we can see in fig. \ref{entro2}. It has a maximum at $t=\frac{\pi}{4\lambda}$, as it should be expected, in view of the behavior of the state (\ref{res1b}).

The maximum entropy is

\begin{equation}
S_{max}= S(t=\frac{\pi}{4\lambda})=\ln d(N), \label{Sm}
\end{equation}

\noindent where $d(N)=2^N$ is the Hilbert space dimension of one theory.  This is consistent with the Bekenstein entropy. It can be shown that  $d(N)$ is also related to the number of modes truly occupied of the two theories in the maximum entanglement condition. Let us consider $N_a(t)$ and $N_b(t)$ to be the expected value of number operator for each theory. So $D_N(t) =2^{N_a(t)}\times 2^{N_b(t)}$ is the dimension of the total occupied Hilbert space. As

\be
N_a=\sum_i^{N}\left\langle 0(t)\right|a^{\dagger}_ia_i\left|0(t)\right\rangle= N\sin^2(\gamma t),\:\:\:
N_b=\sum_i^{N}\left\langle 0(t)\right|b^{\dagger}_ib_i\left|0(t)\right\rangle= N\sin^2(\gamma t)
\ee

\noindent one can see that at the maximum entanglement time, $D_N(t=\frac{\pi}{4\lambda})=2^N=d(N)$. 

\section{Thermal state as initial state}

In the previous analysis the initial state was prepared to be the pure state 
$\left |0\right\rangle = \left|0\right\rangle_A\otimes\left|0\right\rangle_A$. 
In this way, the mechanism that connects the two theories sounds somewhat artificial. However, the same analysis can be done if the initial state is prepared to be a thermal state. That is, an entangled state of the two theories in such a way that, at $t = 0$, we already have a black hole. Let's assume the following initial state

\be
\left |0\right\rangle = \left |0(\beta)\right\rangle=\prod_{n}^{N}e^{G_n(\theta_n)}\left|0\right\rangle_a\otimes\left |0\right\rangle_b
\label{TFD}
\ee 

\noindent where

\begin{equation}
G_n(\theta_n)=\theta_{n}(a_nb_n-a^{\dagger}_nb^{\dagger}_n)
\end{equation}

\noindent is the TFD Bogoliubov generator \cite{ChuUme}. At the equilibrium temperature $T=\frac{1}{\beta}$, the parameter $\theta_n$ is given by

\begin{equation}
\theta_{n}= \arcsin\left(\frac{1}{(e^{\beta E_{n}} +1)^{1/2}}\right).
\label{theta}
\end{equation}

\noindent where  $E_{n}=\langle n|H |n \rangle$ is the energy of one CQM and $Z(\beta)$ is the partition function. Note that the estate (\ref{TFD}) is a ground state of the tunneling Hamiltonian. As $[H_{\lambda},G(\theta_{n})]=0$, the time dependent squeezed vacuum becomes

\begin{eqnarray}
\left|0(\beta,t)\right\rangle_I&=& e^{-itH_{\lambda}}\left |0(\beta)\right\rangle \nonumber \\
&=&\prod_{n}\frac{\delta_{ii}}{\cos\left(\gamma t+\theta_n\right)}\exp\left[\tan\left(\gamma t+\theta_n\right)a^{\dagger}_{n}b^{\dagger}_{n}\right]|0\rangle_A\otimes|0\rangle_B. \label{res1}
\end{eqnarray}

\noindent Although the state seems complicated due to the implicit dependence of $\theta_{n}$ with $\beta E_n$ (given by equation (\ref{theta})), the behavior is similar to that found previously. Now the state starts in a thermal state  (which is a pure state of the total Hilbert space) and at $t = \frac{\pi}{2\lambda}$, it returns to the same state. It can be shown that the state (\ref{res1}) is in fact a time dependent TFD state. It is similar to the one found in \cite{CMD} for a bosonic dissipative system. Note that the state (\ref{res1}) is annihilated by the following operator

\be
a_i(\beta,t)= a_i(t)\cos(\theta_i) -b_i^{\dagger}(t)\sin(\theta_i)  
\ee

\noindent where
 
\be
a_i(t)= a_{i}\cos(t) -b_{i}^{\dagger}\sin(t),\:\:\:\:
b_i^{\dagger}(t)=b_i^{\dagger}\cos(t)-a_i\sin(t)
\ee

\noindent So, the state (\ref{res1}) can be written as

\be
\left|0(\beta,t)\right\rangle&=& \prod_i^Ne^{G_i(\beta,t)}|0\rangle_A\otimes|0\rangle_B.
\ee

\noindent where $G_i(\beta,t)= \theta_i\left(a_i(t)b_i(t) -a_i^{\dagger}(t)b_i^{\dagger}(t)\right)$ is the typical TFD Bogoliubov operator for a system of time dependent fermionic oscillators \cite{KMMS}. The reduced density matrix can be calculated in the same way as in (\ref{dmatrix}):

\begin{eqnarray}
 \rho _a(\beta,t) =T{r_{\{ b\} }} \rho(\beta,t) 
=\prod_{k=1}^{N}\frac{1}{\cos(\gamma t+\theta_k)}\sum_{n_k}\left[\tan^2(\gamma t+\theta_k)\right]^{n_k}|n_k\rangle\langle n_k| 
\end{eqnarray}

\noindent and the entanglement entropy is

\be
S(\beta,t) = -\sum_{n=1}^{N}[\sin^2(\theta_n+\gamma t)\ln \left( \sin^2(\theta_i+\gamma t)\right)
+\cos^2(\theta_n+\gamma t)\ln\left (\cos^2(\theta_n+\gamma t)\right)] \, . \label{timeentropy2}
 \ee

\noindent  Note that, at the limit $e^{-\beta E_{n}}\rightarrow 0$, $\theta_n$ goes to zero and the entropy calculated in (\ref{timeentropy2}) has the same form of (\ref{timeentropy}). Actually, as $E_n$ must be proportional to $J$, this limit corresponds to $\beta J >> 1$ and it is precisely in this limit (for large N) that the SYK model has conformal symmetry and holographic interpretation.

 For the sake of simplicity and to illustrate the results and its dependence on $N$, suppose that $E_n = cn$, where c is a constant; that is, an ensemble of free fermions at $t = 0$.  Now, the Page curve is found only in the large $N$ limit as shown in figure (\ref{entroft}). Notice how the shape of the curve changes as $N$ grows. This is just because in the large $N$ limit  $\theta_n$ goes to zero. At this limit, the maximal entropy is the same as in equation (\ref{Sm}).

\begin{center} 
 \begin{minipage}[t]{0.4\textwidth}
  \includegraphics[width=\textwidth]{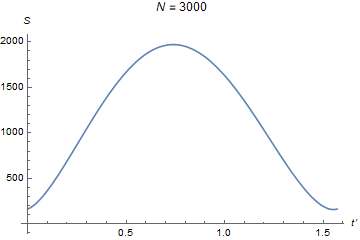} 
	\end{minipage}
	\begin{minipage}[t]{0.4\textwidth}
  \includegraphics[width=\textwidth]{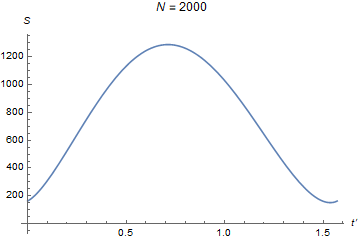}
	\end{minipage}
	
	\begin{minipage}[t]{0.4\textwidth}
  \includegraphics[width=\textwidth]{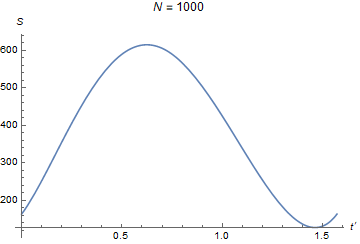}
	\end{minipage}
	\begin{minipage}[t]{0.4\textwidth}
  \includegraphics[width=\textwidth]{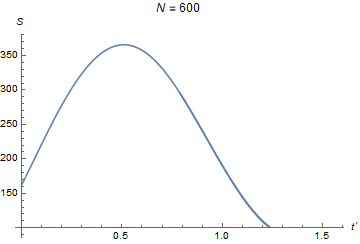}
	\end{minipage}
	
	\begin{minipage}[t]{0.4\textwidth}
  \includegraphics[width=\textwidth]{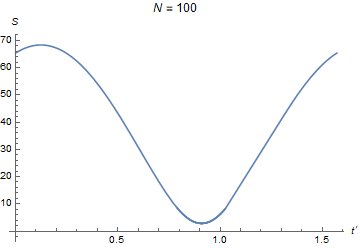}
	\end{minipage}
	\begin{minipage}[t]{0.4\textwidth}
  \includegraphics[width=\textwidth]{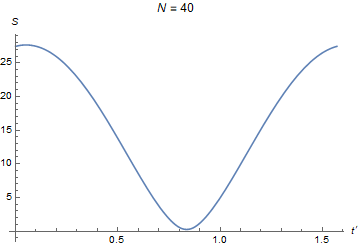}
	\end{minipage} 
\end{center}
\captionof{figure}{The behavior of $S(t)$  as a $t'=\gamma t$ function, from Eq.(\ref{timeentropy}).  The Page curve is acquired at large $N$ limit. It was used $\beta E_n = 0.01n$. }
    \label{entroft}

\section{Entropy operator and modular Hamiltonian}

Finally, an interesting relationship between the approach used here, the TFD entropy operator and the modular Hamiltonian can be found. The TFD state at the infinite temperature limit has the form

\be
\left|0(\beta\rightarrow 0)\right\rangle=\left|I\right\rangle=\prod_{i}e^{a_i^{\dagger}b_i^{\dagger}}\left|0\right\rangle_a\otimes\left|0\right\rangle_a \label{I}
\ee

 \noindent This is, in principle, a non-normalizable state. However, it is possible to introduce a regularization, as shown in \cite{Mollabashi:2013lya}. The state (\ref{res1}) can be obtained from state (\ref{I}) through the following operation

\be
\left|0(\beta,t)\right\rangle =e^{\frac{-K_a(\beta,t)}{2}}\left|I\right\rangle
 \ee

\noindent where
 
\be
K=-\sum^N_{n=1} \, \left(a^\dagger_n \, a_n \, \ln \left(\sin^2 (\theta_n+\gamma t)\right ) + a_n \, a^\dagger_n \, \ln \left(\cos^2 (\theta_n+\gamma t\right) \right)
\ee

\noindent is the entropy operator. Its expected value in state (\ref{res1}) exactly reproduces the entropy calculated at (\ref{timeentropy2}). It appears naturally in dissipative systems \cite{ceravi} and it was defined in the TFD formalism \cite{ChuUme}.  Note that the operator $K(\beta,t)$ leads a system from infinite to finite temperature. The operator also carries on all the time dependency of the state (\ref{res1}) in such a way that

\begin{equation}
\frac{\partial \left|0(\beta,t)\right\rangle }{\partial t}=\frac{-\partial K(\beta,t)}{\partial t}\left|0(\beta,t)\right\rangle
\end{equation}

\noindent so the basic notion of equilibrium, $\displaystyle\frac{\partial\left|0(t)\right\rangle}{\partial t}\approx0$, is equivalent to the maximum entropy condition. The $K(\beta,t)$ operator also has an important relationship with the reduced density matrix. A little algebra shows that

\begin{equation}
\rho_a (\beta,t) = e^{-K(\beta,t)} \ ,
\end{equation}

\noindent which shows that the entropy operator is nothing but the time dependent modular Hamiltonian for this state \cite{haag}.

\section{Conclusion}

In this letter, two modifications of the usual interaction terms between two SYK models was made. The first one is just a Bogoliubov transformation of the fields, changing the bilinear interaction term (known as tunneling Hamiltonian), which is usually studied in this scenario. The second modification is the addition of a new four fermions interaction between the two theories.  This allows to write an interaction picture where the vacuum's time evolution is written in terms of the Bogoliubov transformed tunneling Hamiltonian.  It is shown that the vacuum evolves as a $SU(2)$ entangled state of the two theories. In the first part of this work, the initial state is prepared to be the vacuum of the system. In this case, the system evolves from a pure state at $t = 0$ and returns to the same pure state at $t =\frac{\pi}{2\lambda}$, where $\lambda$ is the coupling constant present in the tunneling Hamiltonian. The entanglement entropy is calculated as a function of time and reproduces the Page curve. The maximal entropy is obtained at $t =\frac{\pi}{4\lambda}$ and is $S_{max}= \ln d(N)$, where $d(N)$ is the dimension of the Hilbert space. The same happens when the initial state is prepared to be a thermal state.  In this case, the Page curve is reproduced only in the large N limit. 

In \cite{Sahoo} it is performed an exact diagonalization of the SYK
model and it is shown that it admits a ground state close to a TFD one. To complete the work presented here, it will be important to study the influence of the new interactions in this diagonalization procedure.  As a future work, it will also be interesting to study the influence of the new interactions proposed  here in the dynamics of the SYK model operators, as well as to calculate the correlation functions in the time dependent entanglement vacuum defined in equation (\ref{res1}) and its subsequent holographic interpretations. 
\begin{acknowledgments}
The author would like to thank Dafni F. Z. Marchioro and M. B. Cantcheff for useful discussions.
\end{acknowledgments}

\end{document}